\documentclass[submission,copyright,creativecommons]{eptcs}
\providecommand{\event}{HSB 2013} 
\usepackage{breakurl}             
\usepackage{amsmath,amsfonts,calrsfs,amssymb,color,graphicx,fixmath,amsthm,longtable}
\title{A hybrid mammalian cell cycle model}
\author{Vincent No\"el
\institute{Universit\'e de Rennes 1}
\email{vincent.noel.pro@gmail.com}
\and Sergey Vakulenko
\institute{Saint Petersburg State University of Technology and Design	}
\email{vakulenfr@mail.ru}
\and Ovidiu Radulescu
\institute{Universit\'e de Montpellier 2}
\email{ovidiu.radulescu@univ-montp2.fr}
}
\def\titlerunning{A hybrid mammalian cell cycle model}
\def\authorrunning{V. No\"el, S. Vakulenko,  \& O. Radulescu}

\newcommand{\vect}[1]{\ensuremath{ \mathbold #1 } }

\begin{document}
\maketitle

\begin{abstract}
Hybrid modeling provides an effective solution to cope with multiple time scales dynamics
in systems biology. Among the applications of this method, one of the most important is the cell cycle regulation. The machinery of the cell cycle, leading to cell division and proliferation, combines slow growth, spatio-temporal re-organisation of the cell, and rapid changes of regulatory proteins concentrations induced by post-translational modifications. The advancement through the cell cycle comprises a well defined sequence of stages, separated by checkpoint transitions. The combination
of continuous and discrete changes justifies hybrid modelling approaches to cell cycle dynamics.
We present a piecewise-smooth version of a mammalian cell cycle model, obtained by hybridization from a smooth biochemical model. The approximate hybridization scheme, leading to simplified reaction rates and binary event location functions, is based on learning from a training set of trajectories of the smooth model. We discuss several learning strategies for the parameters of the hybrid model.
\end{abstract}

\section{Introduction}
Systems biology employs a large number of formalisms to represent the dynamics
of biochemically interacting molecules in signal transduction, metabolic and gene
regulatory networks. Some of these formalisms, such as the systems of ordinary
differential equations (ODE), are based on continuous representations of the
phase space, whereas others, such as boolean networks, employ discrete
dynamical variables. New approaches, based on hybrid models and using
both continuous and discrete variables, are emerging as alternative
descriptions of biochemical networks.

Hybrid modelling allows a good compromise between realistic description of
mechanisms of regulation and the possibility of testing the model in terms
of state reachability and temporal logics \cite{lincoln2004symbolic,mishra2009intelligently}.
Threshold dynamics of gene regulatory networks \cite{baldazzi2011qualitative,ropers2011model}
or of excitable signaling systems \cite{ye2008modelling}
has been modelled by piecewise-linear and piecewise-affine models.
These models have relatively simple structure and can, in certain cases, be
identified from data \cite{porreca2008structural,drulhe2008switching}.
Some methods were proposed for computing the set of reachable states of
piecewise affine models \cite{batt2008symbolic}.

The use of hybrid models in systems biology is justified when some events,
such as rapid protein modifications occur on very short time scales and produce
significant changes of the systems dynamics. The regulatory machinery of the
cell cycle of eukaryotic organisms provides a remarkable example of such a
situation. Indeed, the advancement through the cell cycle consists of a
well defined sequence of stages, separated by checkpoint transitions.
During each one of this stages, different sets of dynamical variables
and biochemical reactions are specifically active, and change from one
stage to another. A hybrid model of mammalian cell cycle has been previously
proposed by Tyson's  group \cite{singhania2011hybrid}. This model is based on a
Boolean automaton whose discrete transitions trigger changes of
kinetic parameters in a set of ODEs. The construction method is ad hoc
and therefore difficult to generalize.  Similar hybrid cell cycle models 
can be found elsewhere \cite{alfieri2011modeling}.

Recently, we have proposed a hybridization method for systematically deriving
hybrid models from smooth ODE models \cite{noel2010,noel2011}.
In this method, non-linear reaction rate functions of biochemical reactions
are approximated by simpler, piecewise linear functions.
The hybrid model contain new parameters that can be estimated 
by a combination of linear programming and least squares optimization.
In this paper we discuss an application of the method to
a medium size cell cycle model.
Our method has some similarities to the method 
proposed in \cite{grosu2007learning} to learn hybrid models from action potentials, 
but there are also differences, such as the definition of the modes and of the mode 
switching, and the optimization scheme.

\section{Piecewise smooth hybrid models}

 We consider piecewise smooth hybrid dynamical systems (HDS) for which the continuous variables, $u$, satisfy the equations
\begin{eqnarray}
  \frac{du_i}{ dt} =  \sum_{k=1}^{N_i} s_{k} P_{ik} (u)+ P_{i}^0 (u) - \sum_{l=1}^{M_i} \tilde s_l Q_{il}(u) - Q_{i}^0 (u), \notag \\
  s_j  = H ( \sum_{k \in C_j} w_{jk} u_k - h_j ), \quad
  \tilde s_l  =  H ( \sum_{k \in \tilde C_l } \tilde w_{lk} u_k - \tilde h_l ),
\label{hybrid}
\end{eqnarray}
where $H$ is the unit step function $H(y)=1, \, y\geq0$, and $H(y)=0, \, y < 0$, $P_{ik}, P_{i}^0, Q_{il}, Q_i^0$ are positive, smooth functions of $u$ representing production, basal
production, consumption, and basal consumption, respectively.
Here $w,\tilde w$ are matrices describing the interactions between the $u$ variables, $i=1,2,...,n$, $j=1,2,..., N$, $l=1,..., M$ and
$h,\tilde h$ are thresholds, and $C_j$, $\tilde C_l$ are indices subsets corresponding to continuous variables controlling
the discrete variables.

One will usually look for solutions of the
 piecewise-smooth dynamics \eqref{hybrid} such that trajectories of $\vect{u}$ are continuous.
  However, we can easily extend the above
definitions in order to cope with jumps of the continuous variables. Similarly to impact systems occurring in mechanics \cite{di2008piecewise}, the jumps of the continuous variables can be commanded by the following rule:
$\vect{u}$ instantly changes to $\vect{p}_j^{\pm}(\vect{u})$ whenever a discrete variable $\hat s_j = H ( \sum_{k \in \hat C_j} \hat w_{jk} u_k - \hat h_j )$ changes. The $\pm$ superscripts
correspond to changes of $\hat s_j$ from
 $0$ to $1$ and from $1$ to $0$, respectively.
 We can consider reversible jumps in which
case  the functions $\vect{p}^{\pm}_j(\vect{u})$ satisfy $\vect{p}^{+} \circ \vect{p}^{-} = Id$.
The typical example
in molecular biology is the cell cycle. In this case, the command to divide
 at the end of mitosis is irreversible and corresponds to $\vect{p}^+_j(\vect{u}) = \vect{u}/2$.
 No return is possible, $\vect{p}^-_j(\vect{u}) = \vect{u}$.

The class of models \eqref{hybrid} is too general. We will restrict ourselves to a subclass
of piecewise smooth systems where
smooth production and degradation terms are assumed multivariate
monomials in $u$, plus some basal terms that we try to make as
simple as possible. A system with constant basal production and linear
basal consumption is the following:
\begin{eqnarray}
P_{ik} (\vect{u}) &= & a_{ik} u_1^{\alpha_1^{ik}} \ldots u_n^{\alpha_n^{ik}} , \notag \\
P_{i}^0 (\vect{u}) & =& a_i^0, \notag \\
 Q_{il} (\vect{u}) & = &\tilde a_{il} u_1^{\tilde \alpha_1^{il}} \ldots u_n^{\tilde \alpha_n^{il}}, \notag \\
Q_{i}^0 (\vect{u}) & = &\tilde a_i^0 u_i.
 \label{rates}
\end{eqnarray}

Multivariate monomial rates represent good approximations for nonlinear networks
of biochemical reactions  with multiple separated timescales \cite{radulescu2008robust,gorban-dynamic}.
Such examples are abundant in chemical kinetics.
For instance, Michaelis Menten, Hill, or Goldbeter-Koshland reactions switch
from a saturated regime where rates are constant to a small concentration regime where
rates follow power laws. The definition of the rates reminds that of S-systems, introduced by
Savageau \cite{savageau1987recasting}. Finally, as discussed in \cite{noel2013} monomial
approximations occur naturally in ``tropically-truncated'' polynomial systems, ie
in systems where polynomial or rational rate functions are replaced by a few dominating monomials.
As compared with our previous work \cite{noel2013,noel2012hybrid}, in this paper we use the tropicalization only
heuristically to obtain simpler reaction kinetic laws, whose parameters are then fitted.
The switching of the monomial terms is not given by the max-plus rule as in \cite{noel2013,noel2012hybrid},
but is commanded by thresholding functions depending on parameters to be fitted.
This allows for more flexibility and corrects the errors introduced by the tropicalization.

\section{Hybridization of the generic mammalian cell cycle model}

This model has been proposed by the group of Tyson \cite{csisz2006analysis} and is designed to be a generic model of the cell cycle for eukaryotes. The cell cycle being an old, but important system that evolved, there have to be homologies, i.e. common mechanisms shared by the cell cycle regulation of all eukaryotes.
The goal of this model is to bring to light these mechanisms, while producing models that reproduce experimental results. Four different eukaryotic organisms were modelled : budding yeast, fission yeast, Xenopus embryos, and mammalian cells. For each of theses organisms, a set of parameters is provided. By changing parameter sets, one can activate or deactivate some modules, fine tune some mechanisms, in order to reproduce the behaviour of the cell cycle in the chosen organism.

We analyse here only the model describing mammalian cells. This model uses twelve variables (eleven of them being concentrations of proteins, and one being the mass of the cell) and forty reactions. We briefly discuss the steps of the algorithm applied to this model.

{\em Choice of the hybrid scheme.}
Five of these reactions  are typically switch-like, following Goldbeter-Koshland kinetics, defined as follows:

\begin{equation}
GK(v_1,v_2,J_1,J_2) = \frac{2v_1J_2}{B+\sqrt{B^2-4(v_2-v_1)v_1J_2}},
\end{equation}
with $B=v_2-v_1+J_1v_2+J_2v_1$.

These kinetics describe a steady-state solution for a 2-state biological system, meaning that this reaction will have two basics modes : active or inactive.
These reactions are replaced by switched reactions
whose rates are simplified monomial rates multiplied by a boolean variable.

For instance the reaction that produces Cyclin-B, induced by the cell mass, has the following kinetic rate:

\begin{equation}
\mathrm{R} = \mathrm{ksb}_{pp}\, \mathrm{[Mass]}\, \mathrm{GK}\!\left(\mathrm{kafb}\, \mathrm{[CycB]},\mathrm{kifb},\mathrm{Jafb},\mathrm{Jifb}\right)
\end{equation}

In this case we replace the Goldbeter-Koshland (GK) function by a step function 
and obtain the following simpler rate:

\begin{small}\begin{equation}
\mathrm{R'} = \mathrm{k'}\, \mathrm{[Mass]}\, \mathrm{s},
\end{equation}\end{small}
where s is a boolean variable.

%


We apply the same method for all the five GK reactions of the model. The original and hybridized
reaction rates can be find in the Table~2.1.

Another set of reactions reactions we want to modify in this model
are Michaelis-Menten (MM) reactions. We want to reproduce the two
functioning modes of Michaelis-Menten kinetics, namely the linear and
the saturated behaviour. The linear behaviour is observed when the substrate
is in low supply. In this case, the flux of the MM reaction will be linear
 with respect to the substrate supply. The saturated behaviour is observed
 when the substrate supply is in excess, and produce a constant flux.
 Our goal is to obtain a hybrid reaction which switches between these two modes, controlled by boolean variables.

There are ten such reactions within this model.
A classic MM reaction rate would be the following :

\begin{equation}
MM(X) = \frac{k.X}{X+k_{m}},
\end{equation}
 and we propose to replace it by the following reaction :

 \begin{equation}
MM(X) = s.k' + \tilde s.k''.X,
\end{equation}
where s is a boolean variable, and $\tilde{s}$ is the complementary of s.

We apply this transformation to all the Michaelis-Menten reactions. The original and hybridized
reaction rates can be found in the Table~2.1.

{\em Detection of the transitions.}
Static event locations follow from the positions of sharp local maxima and minima
of the derivatives of the reactions rates with respect to time (these correspond to
sharp local maxima and minima of the second derivatives of the species concentrations,
with respect to time). We have checked numerically that in the case of
GK functions, these positions are close to the solutions of the equation $v_1 = v_2$.
This property follows from the sigmoidal shape of the GK regulation functions.
It is indeed well known that GK sigmoidal functions have an inflexion point defined
by the condition $v_1 = v_2$, when the activation and inhibition input rates are equal.
If the case of MM functions, these positions are close to the solutions of the
equation $X = k_{m}$.

\begin{figure}
\begin{center}
\includegraphics[width=150mm]{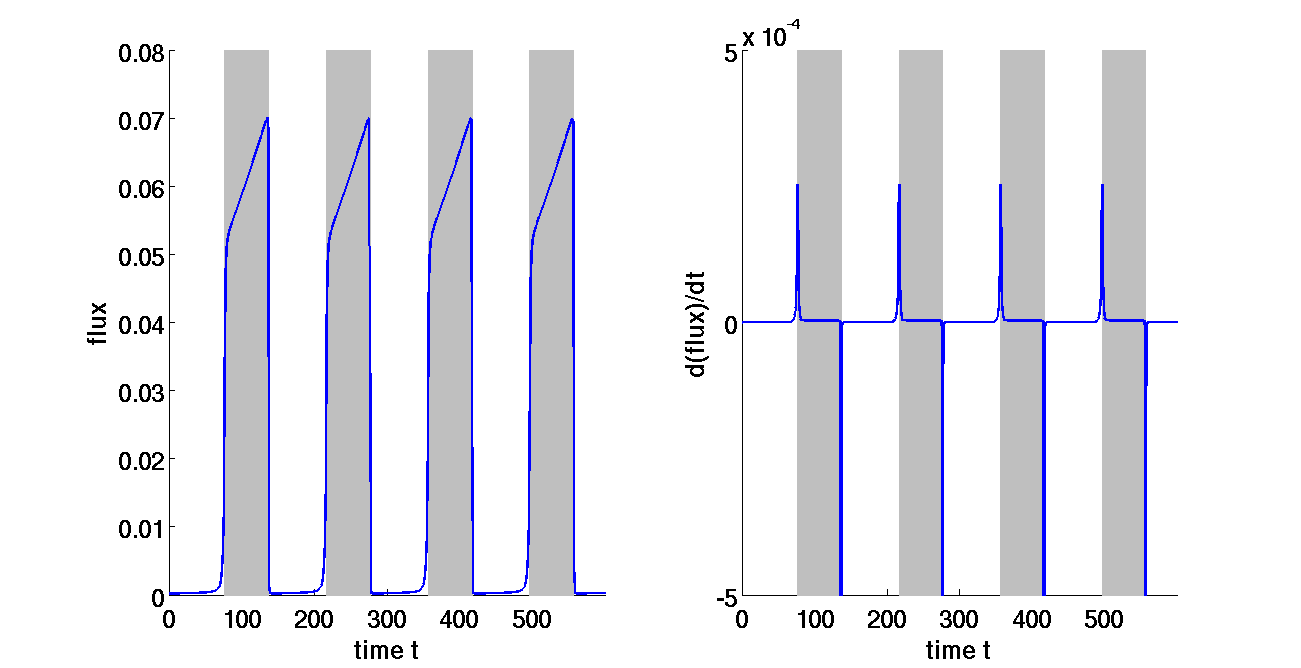}
\caption{Flux and derivative of the flux for the Goldbeter-Koshland reaction R4.
The shaded areas correspond to value where the inequation $v_1 > v_2$ is true.
\label{figdGK}}
\end{center}
\end{figure}

\begin{figure}
\begin{center}
\includegraphics[width=150mm]{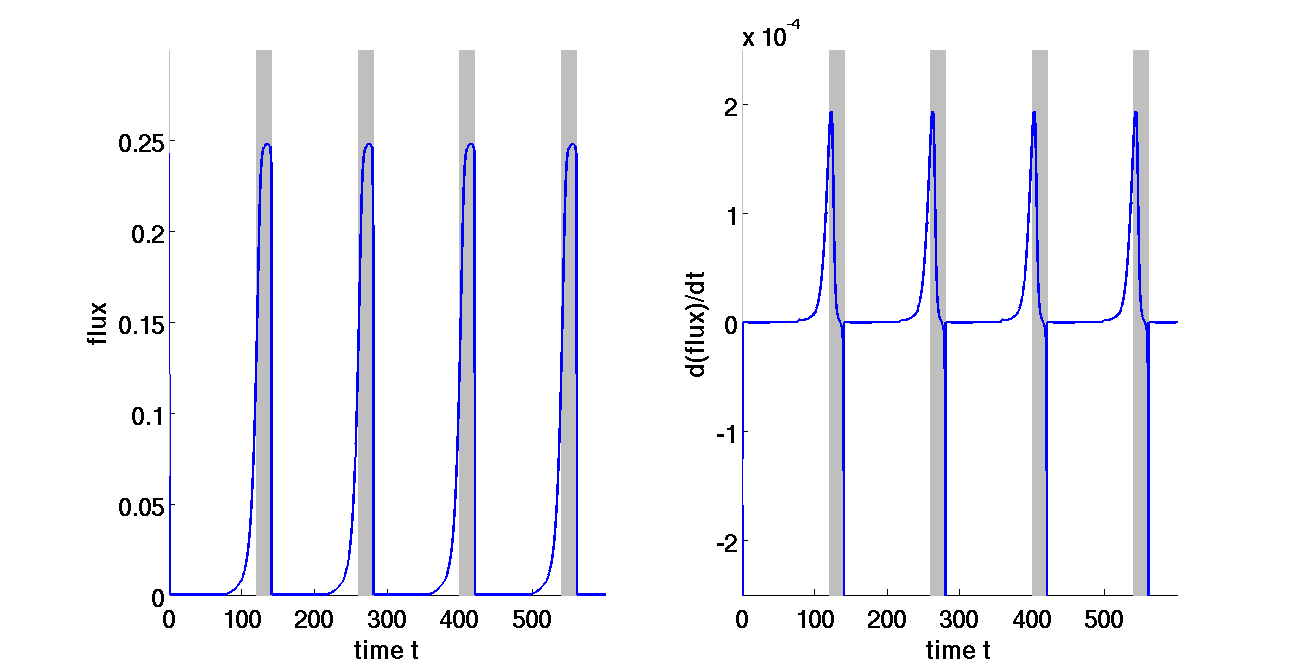}
\caption{Flux and derivative of the flux for the Michaelis-Menten reaction R10. The shaded areas correspond to value where the inequation $X > k_{m}$ is true.
\label{figdMM}}
\end{center}
\end{figure}

These findings are illustrated in Figs. \ref{figdGK},\ref{figdMM}.
We can deduce the value of the boolean variables by checking the inequation $v_1 > v_2$ in the GK case, and $X > k_{m}$ in the MM case. We can observe that the change of the result of these inequations corresponds to the maxima and minima of the derivative (Fig. \ref{figdGK} and Fig. \ref{figdMM}).

The structure of the model can be used to reduce the number of boolean control variables.
In the case of reactions R11,12 or R13,14,15, we can see looking at reaction rates in the Table~2.1 that the inequations controlling their behavior should be the same.
Thus, we can use the same boolean variable to control these reactions.
Furthermore, we found out while looking for these transitions that for some MM reactions
these transitions do not occur along the limit cycle trajectories.
In the case of reactions R7 and R9, the behaviour is always saturated. We chose
not to represent these reactions as hybrid (switched) reactions, and represented
only their saturated behaviour.

We can use these inequalities and hybrid model description to fit parameters
of the hybrid model in one of three ways :

\begin{description}
\item[i)]
Statically, meaning that the discrete variables times series $s(t)$ will be calculated at the detection
step of the algorithm
and will not change
during the fit. In this case one fits only the parameters describing the modes.
This has the benefit of simplicity, but comes with problems. The simplification in the representation of the reactions will introduce a difference between the original and the hybrid model, and such a difference should impact on the position of transitions.
\item[ii)]
Statically, but allow for modifications of the discrete variables time series $s(t)$. We could try to include the positions of these transitions in the fitting parameters, but it would increase the complexity of the cost function. It would notably be a problem to modify all transitions occurring in a single reaction accordingly, which is important for the computation of mode control parameters.
\item[iii)]
Dynamically. We could use the inequations defining the positions of the transitions dynamically, by evaluating them during the optimization. The transitions positions will be determined according to the original model conditions applied to the hybrid model trajectories. This solves the problem of adapting transitions positions of one reaction with respect to the others. The problem is that the transition conditions from the original model are imported to the hybrid model with new parameters (thresholds)
that have also to be fitted.
As thresholds parameters are generally more sensitive, this choice increases the optimization difficulty.
\end{description}

{\em Fitting the hybrid model parameters.}

Once defining the model structure and the parameters to be fitted we can define a cost function
representing the distance between the trajectories of the hybrid and smooth model.
We use a parallel version of Lam's simulated annealing
algorithm \cite{lam1988efficient,reinitzparallel} to minimize this cost function with respect to the parameters of the hybrid model.
 We limit the parameters search space to those involved in the hybridized reactions (a more extensive search is nevertheless possible).
For the cost function, we have decided to test both species trajectories and reaction fluxes. When we limit ourselves to species trajectories, since some reactions have transitions that are close in time, there is a risk that some hybridized reaction will compensate for others. We wanted each hybridized reaction to be as much as possible a replica of the original reaction.

\begin{figure}[h!]
\begin{center}
\includegraphics[width=150mm]{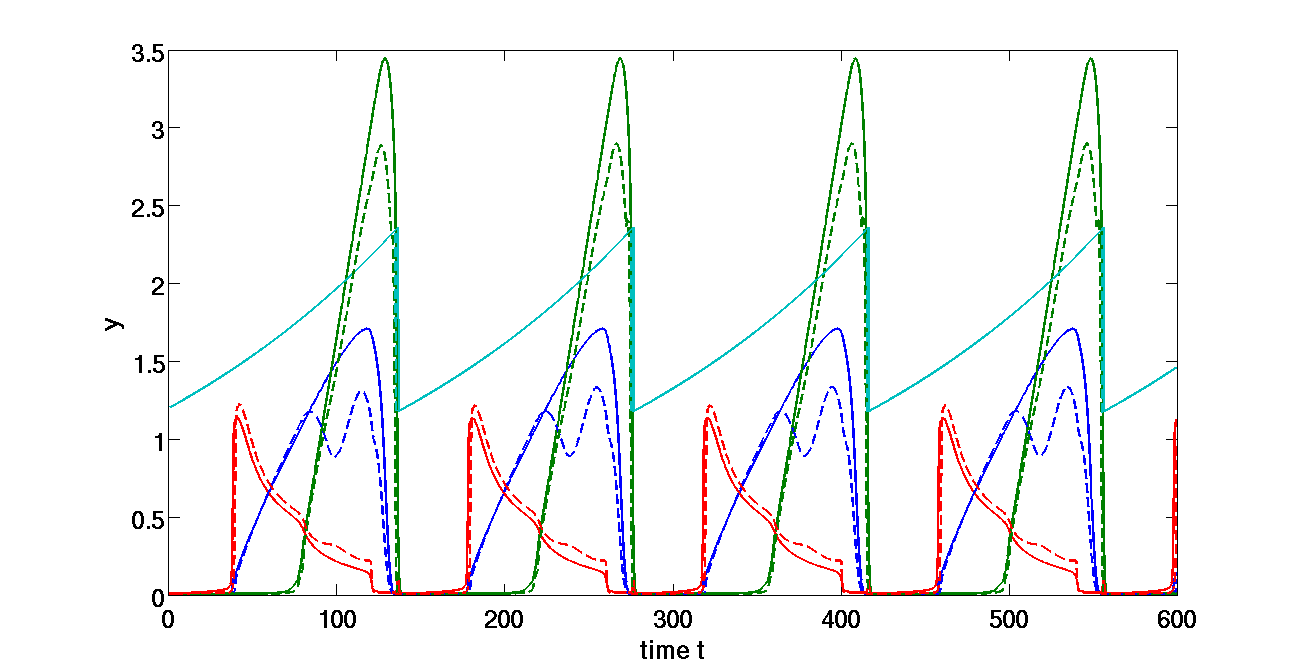}
\end{center}
\caption{Comparison of the trajectories of the four main variables. (blue:~Cyclin-A, green:~Cyclin-B,
red:~Cyclin-C, aqua:~cell size) (Plain lines) Original model (Dashed lines) Hybrid model without mode control parameters fitting (case ii)).
\label{fig_gcc_comp_step0}}
\end{figure}

When using the definition with static discrete variables
 $s(t)$ and fitting only the mode parameters (cases i) above), we were not able to obtain
 even an imperfect fit of the model (in this case the trajectories of the hybrid model
 are very different from the ones of the original model and even become instable). We
 chose to include transition positions to the parameters of the
 fitting (case ii)), and were able to obtain a reasonable fit.
However, the imperfections in the localisation of these new transition positions made difficult to
find good control parameters (see next step) for all the hybridized reactions. The trajectories of the hybrid model fitted using this method are shown in Fig. \ref{fig_gcc_comp_step0}. One can notice important differences between the trajectories of the hybrid and  original model, although these differences remain bounded and the stability of the
limit cycle oscillations is preserved.

\begin{figure}[h!]
\begin{center}
\includegraphics[width=150mm]{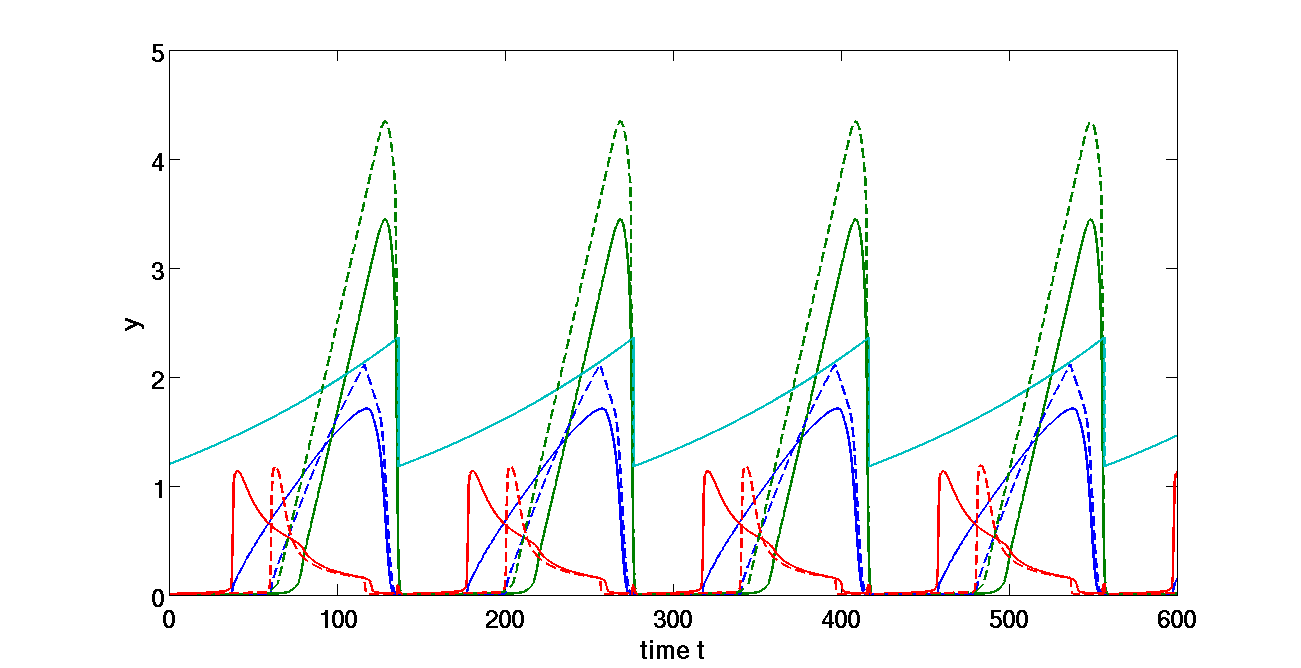}
\end{center}
\caption{Comparison of the trajectories of the four main variables. (blue:~Cyclin-A, green:~Cyclin-B,
red:~Cyclin-C, aqua:~cell size) (Plain lines) Original model (Dashed lines) Hybrid model without mode control parameters fitting (case iii)).
\label{fig_gcc_comp_step1}}
\end{figure}

When using the definition with the original model conditions for transitions (case iii)), we were able to obtain a working hybrid model, but the fit can still be improved by modifying slightly the mode control parameters. We can observe on Fig. \ref{fig_gcc_comp_step1} that while the dynamics of the model is preserved, there are differences in the transition positions.

\begin{figure}[h!]
\begin{center}
\includegraphics[width=150mm]{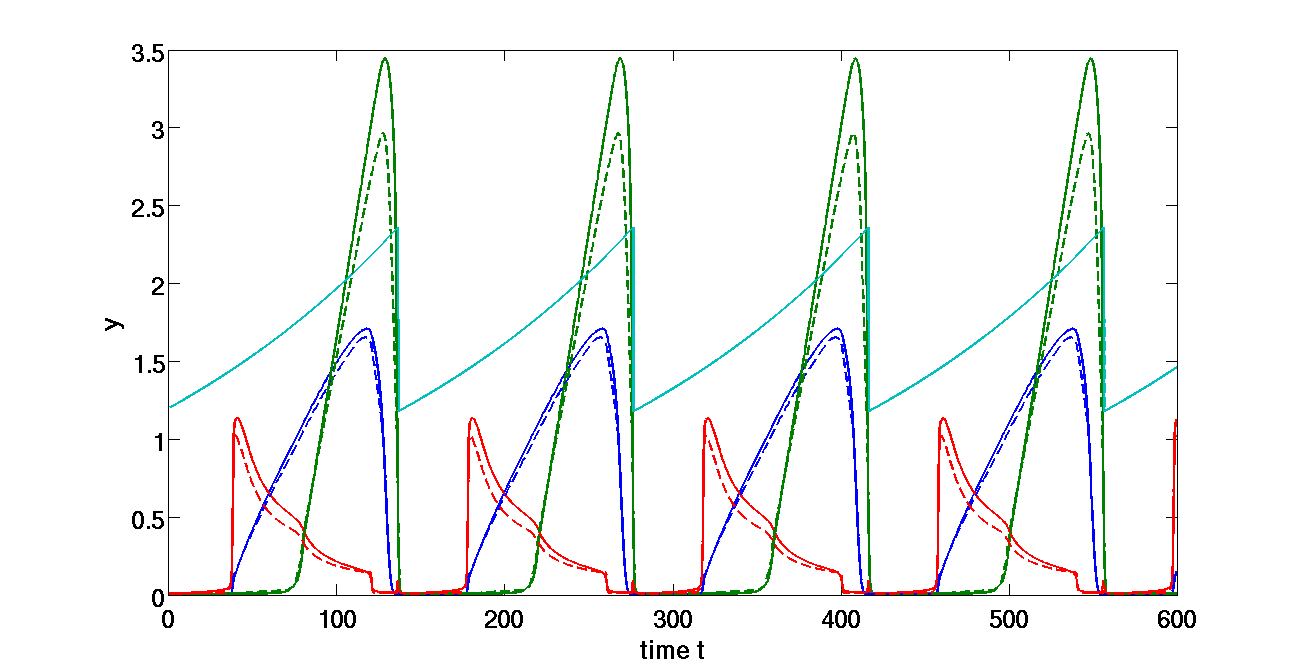}
\end{center}
\caption{Comparison of the trajectories of the four main variables. (blue:~Cyclin-A, green:~Cyclin-B,
red:~Cyclin-C, aqua:~cell size) (Plain lines) Original model (Dashed lines) Hybrid model with mode control parameters fitting (case iii))W.
\label{fig_gcc_comp}}
\end{figure}

\begin{figure}[h!]
\begin{center}
\includegraphics[width=75mm]{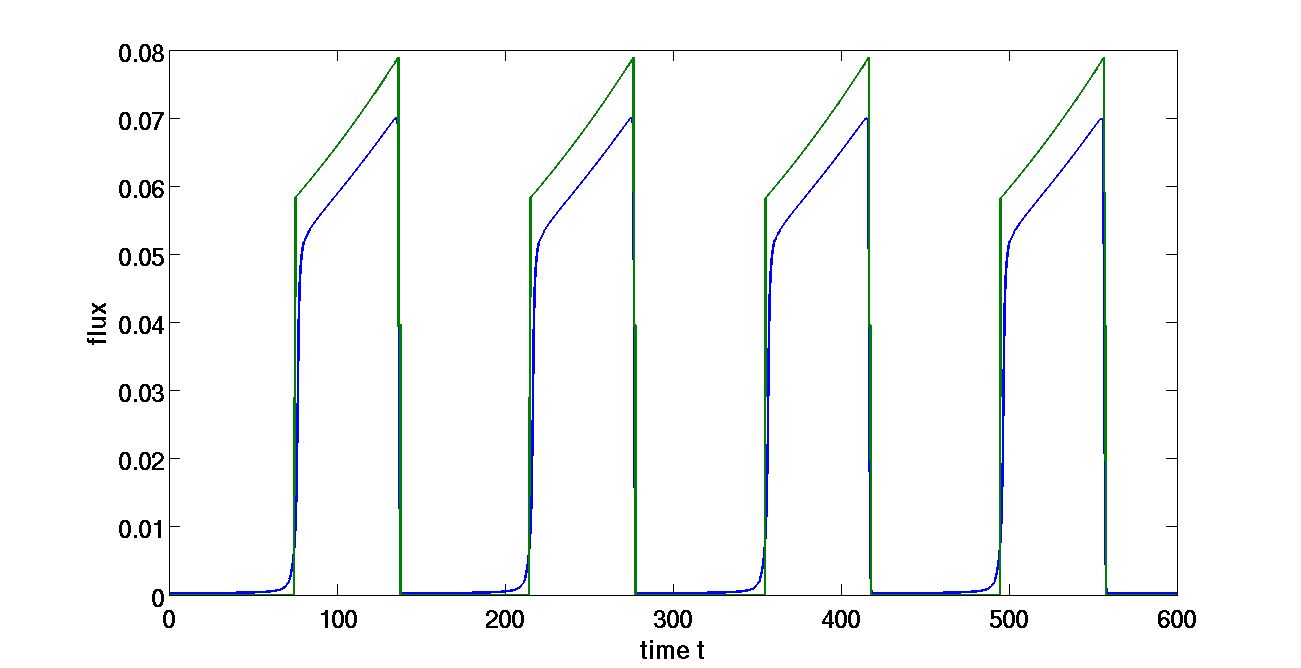}
\includegraphics[width=75mm]{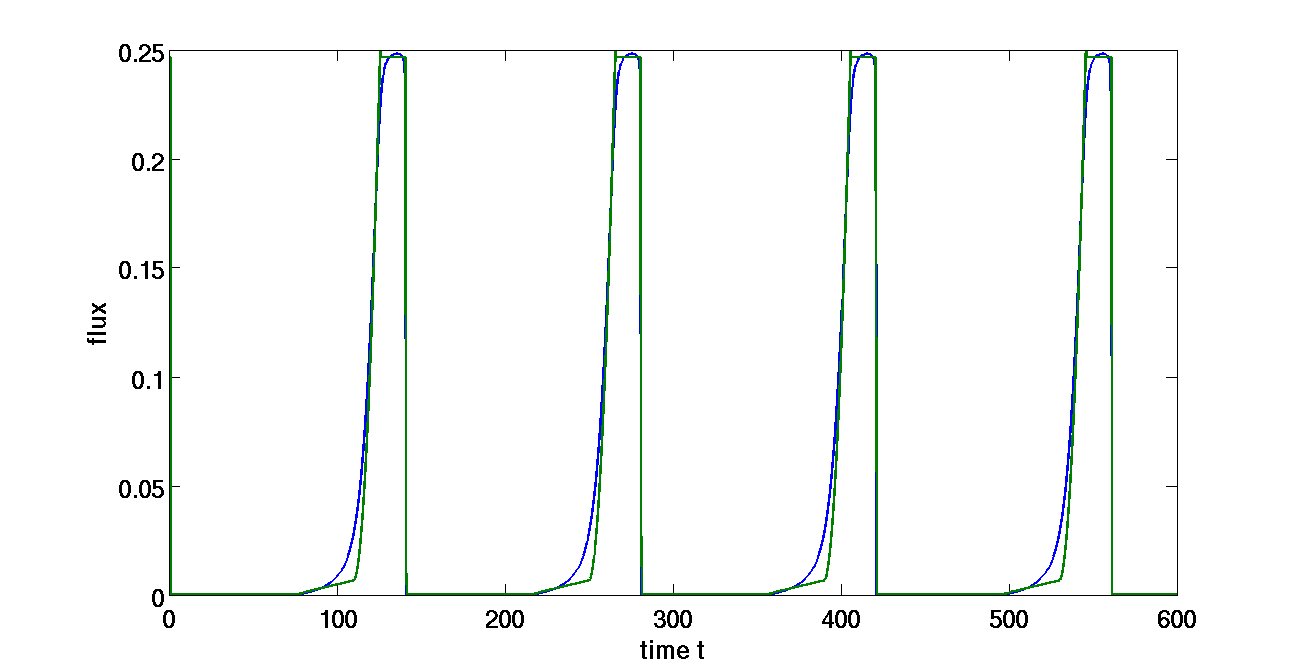}
\caption{Comparison of original and hybridized reaction fluxes. Top : GK Reaction R4. Bottom : MM Reaction R10. Blue : flux of original reaction, Green : flux of hybridised reaction \label{fig_gcc_comp_fluxes}}
\end{center}
\end{figure}

\begin{figure}[h!]
\begin{center}
\includegraphics[width=75mm]{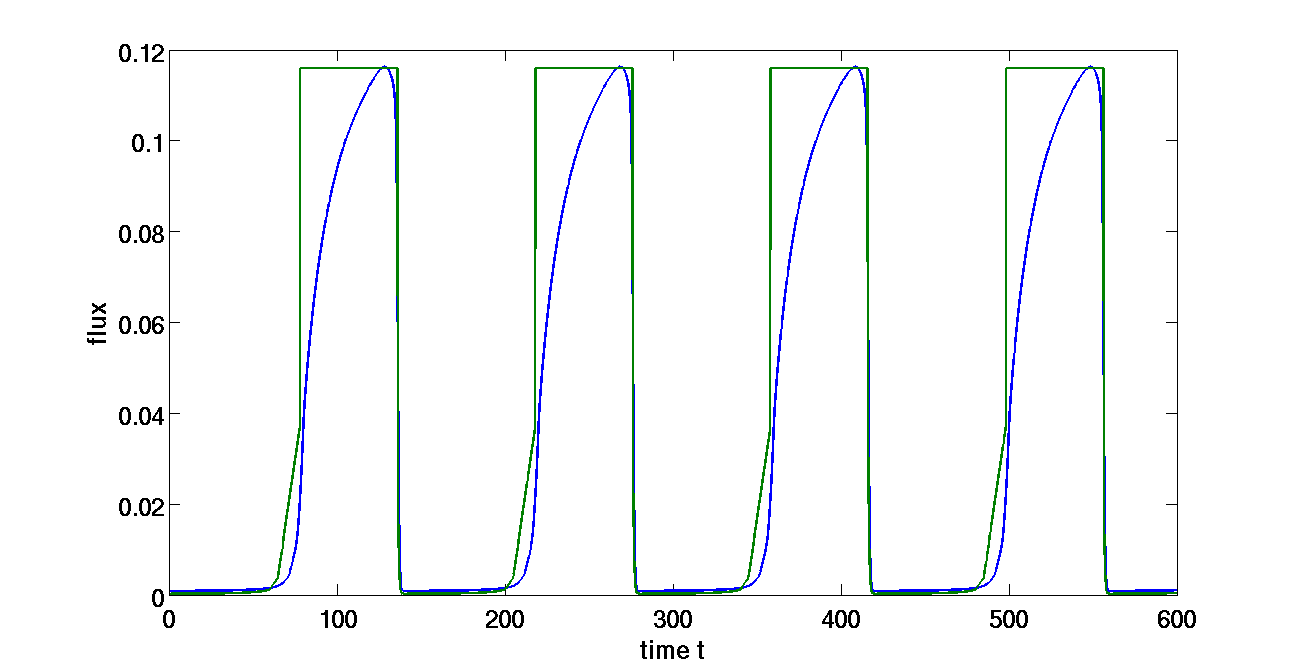}
\includegraphics[width=75mm]{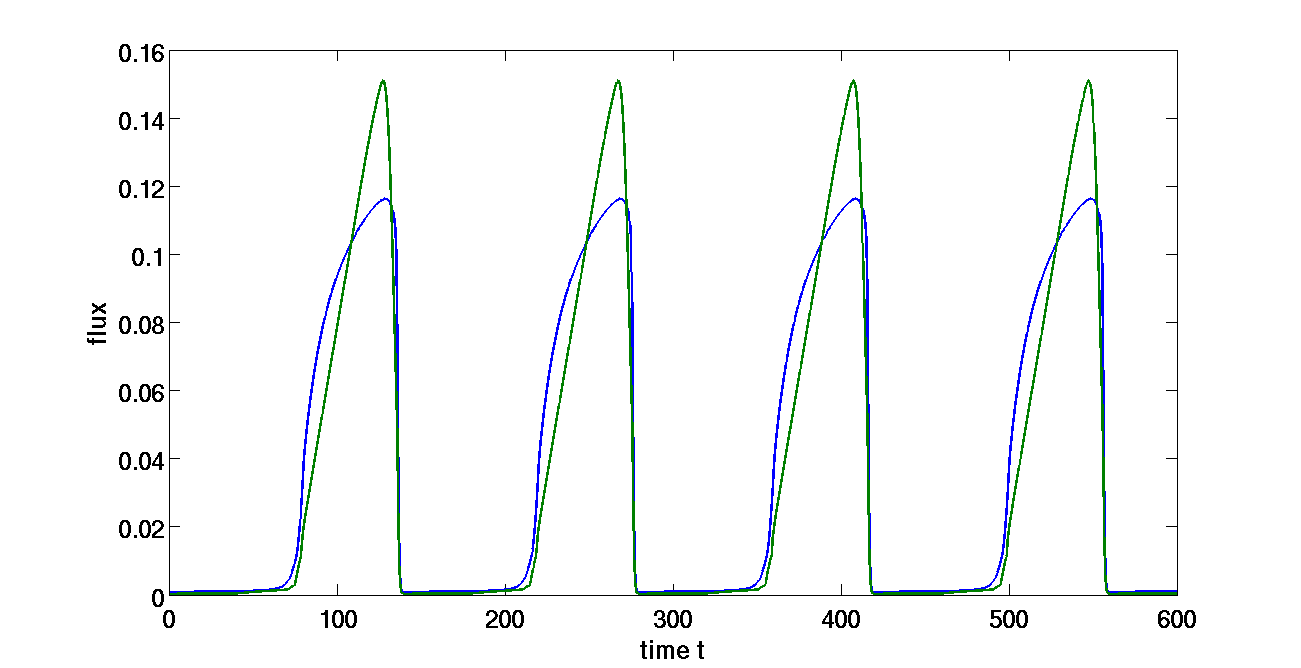}
\end{center}
\caption{Comparison of reaction R6 results without and with the fitting of transition control parameters. Top : Fit without control parameters. Bottom : Fit with control parameters. Blue : flux of original reaction, Green : flux of hybridized reaction
\label{fig_r6}}
\end{figure}

Thus, when we included the parameters of transitions conditions, we obtained a model which fits better the original one. As a control we can see the results of the fitting on both the trajectories of the four main variables (Fig.\ref{fig_gcc_comp}) and the fluxes of some hybridized reactions (Fig.\ref{fig_gcc_comp_fluxes}). \\
An interesting result of this optimization is that some hybridized reactions stopped having transitions, suggesting that the best fit would be obtained without these transitions. The reaction R6 (Fig. \ref{fig_r6}) is one of these reactions. This could be the result of the sensitivity of transition control parameters and a selection of a more robust solution.

%

{\em Computing the mode control parameters.}
If we chose the static method of representing transitions during the fit, we now have to determine a regulation matrix, which will allow a dynamic definition of the events location.

Let $s_m = H ( \sum_{j \in C_m} w_{mj} u_j - h_j)$ be the discrete variables and
 $s_k^m$ the constant values of $s_m$ piecewisely, on time intervals $I_{k}$ identified
at the detection step.
Consider now the optimal trajectories $u_i^{*}(t_l)$, calculated piecewisely with fitted mode parameters.

Then, one should have
\begin{equation}
(\sum_{j \in C_m} w_{mj} u_j^{*}(t_l)  - h_j)  s_k^m > 0, \text{for all} \, t_l \in I_{k}.
\label{mode_control_inequation}
\end{equation}
This is a problem of linear programming, and is solved using the simplex algorithm \cite{dantzig1955generalized}.

At this step, it is interesting to note that we have some choice on which variable can control a given reaction, i.e. on the
subsets $C_m$. This potentially leads to multiple solutions of the inequations. The best choice would be here to use the biological knowledge to choose the species actually involved in the reaction.

The problem with this step is that its success depends on the conservation of the 
transition positions between the simulations with static and dynamic mode control. Or, this
property is valid only to some extent and the dynamic transitions can shift with respect
to their static positions. As a consequence, solving all the 
inequations \eqref{mode_control_inequation} may sometimes be impossible.

To cope with this issue, we introduced a variable $\epsilon$ so that the inequalities \eqref{mode_control_inequation} are modified to :

\begin{equation}
(\sum_{j \in C_m} w_{mj} u_j^{*}(t_l)  - h_j)s_k^m + \epsilon > 0, \text{for all } \quad t_l \in I_{k},
\end{equation}

This modification enables us to solve all the inequations, and gives us a good metric to asses the quality of the resolution. Furthermore, the parameter $\epsilon$ can be minimized within the simplex algorithm. The ideal case is when $\epsilon$ is negative or zero.
When simulating the hybrid model, we found out that with a positive epsilon, the model is most of the time unstable.

\begin{figure}[h!]
\begin{center}
\includegraphics[width=150mm]{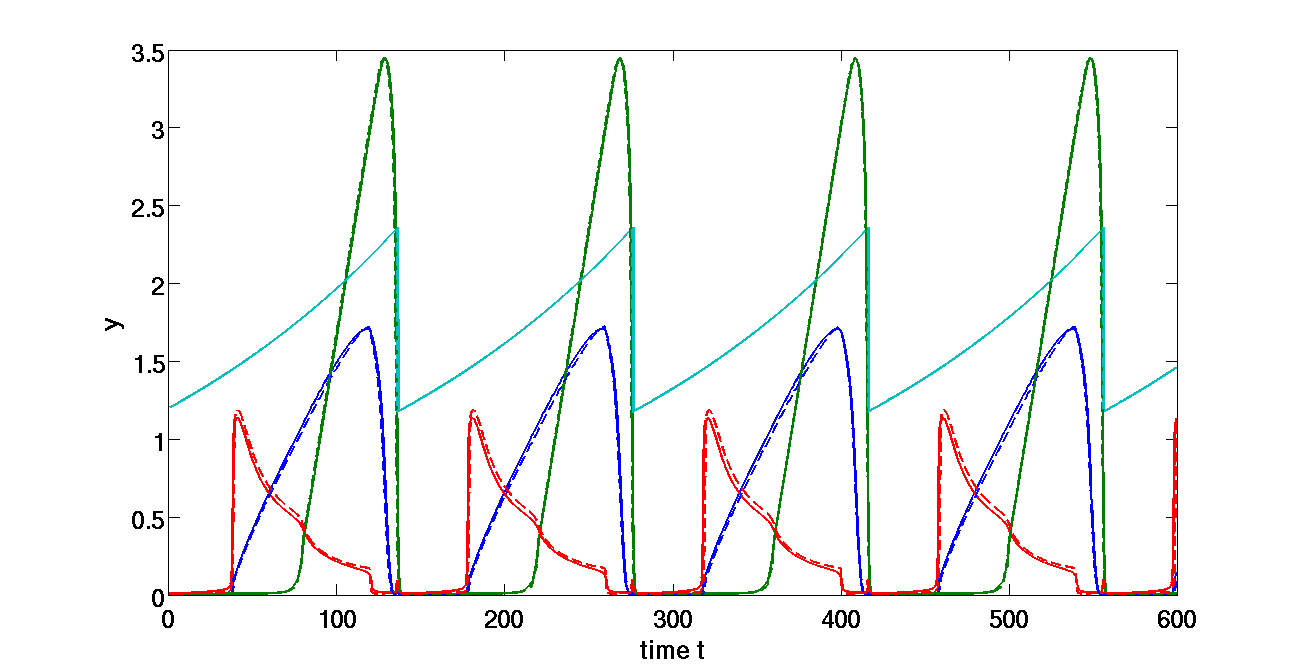}
\end{center}
\caption{Comparison of the trajectories of the four main variables. (blue:~Cyclin-A, green:~Cyclin-B,
red:~Cyclin-C, aqua:~cell size) (Plain lines) Original model (Dashed lines) Hybrid model with mode control parameters fitting.
\label{fig_gcc_comp_4r}}
\end{figure}

Periodicity is not the only difficulty for this step. In our formalism, the threshold to modify the boolean variables controlling a given reaction is the same for an activation or an inactivation. This could also be a problem, as we can not always enforce such a condition during the fitting.
There are different solutions to this problem. The first one would be to have different thresholds for reaction activation and inactivation, but this choice misses the simplicity of the previous method of control. More precisely, even activation and inactivation thresholds correspond geometrically to
control of the modes by manifold crossing (activation when crossing takes place in one direction, inactivation
for crossing in the opposite direction), whereas different thresholds do not allow for such a simple picture.
\\
The other solution would be not to limit ourselves to the biologically relevant variables to control these transitions. As we increase the number of variables, the probability to find a combination which satisfies the inequations increases. The problem with this choice is the large number of possible combinations. We used a genetic algorithm which selects the variables which had the lowest $\epsilon$ value and were able to find combinations which satisfy the inequations for some reactions. But for others reaction, especially Michaelis-Menten reactions, even with all variables, we were not able to obtain low enough $\epsilon$.
We were able to use this method to build a hybrid model which only hybridized the Goldbeter-Koshland reactions. The result can be seen in Fig. \ref{fig_gcc_comp_4r} and the corresponding model is given
in Table 2.3.

\section{Conclusion}

We have presented a hybridization scheme, allowing to transform a biochemical network model,
containing reactions with complex non-linear rate functions, into a hybrid model with piece-wise linear rate functions. This scheme can be applied  to any model containing Goldbeter-Koshland or Michaelis-Menten kinetic laws. More generally, extensions of this method can be applied
to biochemical network models whose kinetic laws are rational functions of the species
concentrations. These include the Goldbeter-Koshland case, as this mechanism is obtained by
model reduction from two coupled Michaelis-Menten reactions. The resulting hybrid model
in the general case is piecewise smooth, but not necessarily linearly smooth.
This generalization is based on tropicalization \cite{SASB2011,noel2013} and consists
in approximating rational rate functions by tropical polynomials, that are represented
piecewisely by multivariate monomials.

The identification algorithm proposed in the paper combines the static or dynamic location of the events, the identification of the mode parameters by simulated annealing, and the identification of the mode control parameters by linear programming. The hardest step of this algorithm
is the simulated annealing. We have discussed three optimization strategies to reduce the number of the
parameters to be determined by simulated annealing, while keeping the flexibility of the optimization
scheme.  In this paper, the hybrid cell cycle model has been obtained from artificial trajectories
generated with a smooth model. That allowed us to include both concentration and rates trajectories
in the cost function, which is a strong constraint. In the future, this constraint could be released
and cost functions based on concentration trajectories only, could be used to learn hybrid
cell cycle models directly from experimental data.

\section{Acknowledgements}
VN thanks University of Rennes 1 for supporting his research and CNRS PEPS MODREDBIO for sponsoring his participation to HSB2013.

\section*{Tables}

\subsection*{Table 2.1 - Definition of reactions in the original
and hybridized mammalian cell cycle model. The inequalities controlling
the mode switching result directly from the definition of the
reaction rates in the original model.}

{\small
\begin{longtable}{  p{0.199\linewidth}  p{0.3\linewidth}  p{0.25\linewidth}  p{0.15\linewidth} }
\hline
reaction smooth & variables & reaction hybrid & control  \\
\hline \\
$R_1 = ksa_{pp}.[Mass].$  &
$v_1=katf_p + katfa_{pp}.[CycA] +$ &
$R_{1h} =ksa_{pp}.[Mass].s_1$ &
$s_1 = v_1 > v_2$
\\
$.GK(v_1,v_2,Jatf,Jitf)$  &
$+ katfd_{pp}.CycD0.[Mass]$
&
&
\\
&
$v_2=kitf_p + kitfa_{pp}.$
&
&
\\
&
$.[CycA]+kitfb_{pp}.[CycB]$
&
&
\\

\hline
$R_2 = k25_{pp}.[pB]$ &
$v_1 = ka25_p+ka25_{pp}.[CycE]$ &
$R_{2h} = k25_{pp}.[pB].s_2$ &
$s_2 = v_1 > v_2$
\\
$.GK(v_1,v_2,Ja25,Ji25)$ &
$v_2 = ki25_p+ki25_{pp}.[Cdc20A]$ &
&
\\

\hline
$R_3 = kwee_{pp}.[CycB]$ &
$v_1 = kawee_p+kawee_{pp}.[Cdc20A]$ &
$R_{3h} = kwee_{pp}.[CycB].s_3$ &
$s_3 = v_1 > v_2$
\\
$.GK(v_1,v_2,Jawee,Jiwee)$ &
$v_2 = kiwee_p+kiwee_{pp}.[CycB]$ &
&
\\

\hline
$R_4 = ksb_{pp}.[Mass]$ &
$v_1 = kafb.[CycB]$ &
$R_{4h} = ksb_{pp}.[Mass].s_4$ &
$s_4 = v_1 > v_2$
\\
$.GK(v_1,v_2,Jafb,Jifb)$ &
$v_2 = kifb$ &
&
\\

\hline
$R_5 = kse_{pp}.[Mass]$ &
$v_1 = katf_p + katfa_{pp}.[CycA] +$ &
$R_{5h} = kse_{pp}.[Mass].s_1$ &
$s_1 = v_1 > v_2$
\\
$.GK(v_1,v_2,Jatf,Jitf)$ &
$+ katfd_{pp}.CycD0.[Mass]$ &
&
\\
&
$v_2=kitf_p + kitfa_{pp}.$ &
&
\\
&
$.[CycA]+kitfb_{pp}.[CycB]$ &
&
\\

\hline
$R_6 = ks20_{pp}$ &
$X = [CycB]$&
$R_{6h} = ks20_{pp}.s_5 $ &
$s_5 = X > K_m$
\\
$.X/(K_m+X)$ &
$K_m = J20$&
$+ ks20_{pp2}.\tilde s5.X$ &
\\

\hline
$R_7 = kaie.[CycB]$ &
$X = (APCT - [APCP])$&
$R_{7h} = ks20_{pp}.[CycB] $ &
\\
$.X/(K_m+X)$ &
$K_m = Jaie$&
&
\\

\hline
$R_8 = kiie$ &
$X = [APCP]$&
$R_{8h} = kiie.s_6$ &
$s_6 = X > K_m$
\\
$.X/(K_m+X)$ &
$K_m = Jiie$&
$+ kiie2*\tilde s_6.X$ &
\\

\hline
$R_9 = ka20.[APCP]$ &
$X = [Cdc20i]$&
$R_{9h} = ka20.[APCP]$ &
\\
$.X/(K_m+X)$ &
$K_m = Ja20$&
&
\\

\hline
$R_{10} = ki20$ &
$X = [Cdc20A]$&
$R_{10h} = ki20.s_7$ &
$s_7 = X > K_m$
\\
$.X/(K_m+X)$ &
$K_m = Ji20$&
$+ ki202.\tilde s_7.X $ &
\\

\hline
$R_{11} = kah1_{p}$ &
$X = (Cdh1T - [Cdh1])$&
$R_{11h} = kah1_{p}.s_8 $ &
$s_8 = X > K_m$
\\
$.X/(K_m+X)$ &
$K_m = Jah1$&
$+ kah1_{p2}.\tilde s_8.X $ &
\\

\hline
$R_{12} = kah1_{pp}.[Cdc20A]$ &
$X = (Cdh1T - [Cdh1])$&
$R_{12h} = kah1_{pp}.[Cdc20A].s_8 $ &
$s_8 = X > K_m$
\\
$.X/(K_m+X)$ &
$K_m = Jah1$&
$+ kah1_{pp2}.[Cdc20A].\tilde s_8.X$ &
\\
\hline
$R_{13} = kih1a_{pp}.[CycA]$ &
$X = [Cdh1]$ &
$R_{13h} = kih1a_{pp}.[CycA].s_9$ &
$s_9 = X > K_m$
\\
$.X/(K_m+X)$ &
$K_m = Jih1$&
$+ kih1a_{pp2}.[CycA].\tilde s_9.X$ &
\\
\hline
$R_{14} = kih1b_{pp}.[CycB]$ &
$X = [Cdh1]$ &
$R_{14h} = kih1b_{pp}.[CycB].s_9$ &
$s_9 = X > K_m$
\\
$.X/(K_m+X)$ &
$K_m = Jih1$&
$+ kih1b_{pp}.[CycB].\tilde s_9.X$ &
\\
\hline
$R_{15} = kih1e_{pp}.[CycE]$ &
$X = [Cdh1]$ &
$R_{15h} = kih1e_{pp}.[CycE].s_9$ &
$s_9 = X > K_m$
\\
$.X/(K_m+X)$ &
$K_m = Jih1$&
$+ kih1e_{pp}.[CycE].\tilde s_9.X$ &

\\
\hline
\end{longtable}
}

\subsection*{Table 2.2.1 - Parameters of the original mammalian cell cycle
model described in the table~2.1.}

{\small
\begin{longtable}{  p{0.45\linewidth}  p{0.5\linewidth}  }
\hline
constant & value   \\
\hline \\
$kse_{pp}$ & 0.18\\
\hline
$katfa_{pp}$ & 0.2\\
\hline
$katfd_{pp}$ & 3.0\\
\hline
$katfe_{pp}$ & 0.5\\
\hline
$kitf_{p}$ & 0.25\\
\hline
$kitfa_{pp}$ & 0.1\\
\hline
$kitfb_{pp}$ & 0.1\\
\hline
$ksb_{pp}$ & 0.03\\
\hline
$kwee_{pp}$ & 0.2\\
\hline
$k25_{pp}$ & 5\\
\hline
$kafb$ & 1.0\\
\hline
$kifb$ & 0.1\\
\hline
$ksa_{pp}$ & 0.025\\
\hline
$kaie$ & 0.07\\
\hline
$kiie$ & 0.18\\
\hline
$Jaie$ & 0.01\\
\hline
$Jiie$ & 0.01\\
\hline
$ks20_{pp}$ & 0.15\\
\hline
$J20$ & 1\\
\hline
$ka20$ & 0.5\\
\hline
$ki20$ & 0.25\\
\hline
$Ji20$ & 0.0050\\
\hline
$kah1_{p}$ & 0.18\\
\hline
$kah1_{pp}$ & 3.5\\
\hline
$kih1a_{pp}$ & 0.2\\
\hline
$kih1b_{pp}$ & 1.0\\
\hline
$kih1e_{pp}$ & 0.1\\
\hline
$Jah1$ & 0.01\\
\hline
$Jih1$ & 0.01\\
\hline
$kawee_{p}$ & 0.3\\
\hline
$kiwee_{pp}$ & 1.0\\
\hline
$ka25_{pp}$ & 1\\
\hline
$ki25_{p}$ & 0.3\\
\hline

\end{longtable}
}

\subsection*{Table 2.2.2 - Parameters of the hybridized mammalian cell cycle
model described in the table~2.1.}

{\small
\begin{longtable}{  p{0.45\linewidth}  p{0.5\linewidth}  }
\hline
constant & value   \\
\hline \\
$ksa_{pp}$ & 0.024635\\
\hline
$katf_{p}$ & 0\\
\hline
$katfa_{pp}$ & 0.00090318\\
\hline
$katfd_{pp}$ & 2.6897\\
\hline
$katfe_{pp}$ & 2.1407\\
\hline
$kitf_p$ & 0.22282\\
\hline
$kitfa_{pp}$ & 0\\
\hline
$kitfb_{pp}$ & 0.14253\\
\hline
$k25_{pp}$ & 3.559\\
\hline
$ka25_{p}$ & 0\\
\hline
$ka25_{pp}$ & 21.93\\
\hline
$ki25_p$ & 5.425\\
\hline
$ki25_{pp}$ & 0\\
\hline
$kwee_{pp}$ & 0.096009\\
\hline
$kawee_{p}$ & 3.5714\\
\hline
$kawee_{pp}$ & 0\\
\hline
$kiwee_{p}$ & 0\\
\hline
$kiwee_{pp}$ & 9.003\\
\hline
$ksb_{pp}$ & 0.033299\\
\hline
$kafb$ & 0.15998\\
\hline
$kifb$ & 0.0056319\\
\hline
$kse_{pp}$ & 0.14842\\
\hline
$ks20_{pp}$ & 0\\
\hline
$ks20_{pp2}$ & 0.048074\\
\hline
$J20$ & 3\\
\hline
$kaie$ & 0.076693\\
\hline
$kiie$ & 0.1685\\
\hline
$kiie2$ & 17.568\\
\hline
$Jiie$ & 0.0096156\\
\hline
$ka20$ & 0.4815\\
\hline
$ki20$ & 0.24271\\
\hline
$ki202$ & 5.3118\\
\hline
$Ji20$ & 0.045084\\
\hline
$kah1_{p}$ & 0\\
\hline
$kah1_{p2}$ & 0.15387\\
\hline
$kah1_{pp}$ & 0\\
\hline
$kah1_{pp2}$ & 3.9768\\
\hline
$Jah1$ & 1\\
\hline
$kih1a_{pp}$ & 0.099851\\
\hline
$kih1a_{pp2}$ & 2.5689\\
\hline
$kih1b_{pp}$ & 0\\
\hline
$kih1b_{pp2}$ & 14.966\\
\hline
$kih1e_{pp}$ & 0.10502\\
\hline
$kih1e_{pp2}$ & 1.7755\\
\hline
$Jih1$ & 0.12035\\

\hline
\end{longtable}
}

\subsection*{Table 2.3 - Definition of reactions in the original
and hybridized mammalian cell cycle model. The inequalities controlling
the mode switching result from the computation of mode control parameters post-fitting.}

{\small
\begin{longtable}{  p{0.222\linewidth}  p{0.35\linewidth}  p{0.35\linewidth} }
\hline
reaction smooth & variables & reaction hybrid  \\
\hline \\
$R_1 = ksa_{pp}.[Mass].$  &
$v_1=katf_p + katfa_{pp}.[CycA] +$ &
$R_{1h} =ksa_{pp}.[Mass].s_1$
\\
$.GK(v_1,v_2,Jatf,Jitf)$  &
$+ katfd_{pp}.CycD0.[Mass]$ &
\\
&
$v_2=kitf_p + kitfa_{pp}.$ &
\\
&
$.[CycA]+kitfb_{pp}.[CycB]$ &
\\
&
&
\\
\hline
$R_2 = k25_{pp}.[pB]$ &
$v_1 = ka25_p+ka25_{pp}.[CycE]$ &
$R_{2h} = k25_{pp}.[pB].s_2$
\\
$.GK(v_1,v_2,Ja25,Ji25)$ &
$v_2 = ki25_p+ki25_{pp}.[Cdc20A]$ &
\\

\hline
$R_3 = kwee_{pp}.[CycB]$ &
$v_1 = kawee_p+kawee_{pp}.[Cdc20A]$ &
$R_{3h} = kwee_{pp}.[CycB].s_3$
\\
$.GK(v_1,v_2,Jawee,Jiwee)$ &
$v_2 = kiwee_p+kiwee_{pp}.[CycB]$ &
\\

\hline
$R_4 = ksb_{pp}.[Mass]$ &
$v_1 = kafb.[CycB]$ &
$R_{4h} = ksb_{pp}.[Mass].s_4$
\\
$.GK(v_1,v_2,Jafb,Jifb)$ &
$v_2 = kifb$ &

\\

\hline
$R_5 = kse_{pp}.[Mass]$ &
$v_1 = katf_p + katfa_{pp}.[CycA] +$ &
$R_{5h} = kse_{pp}.[Mass].s_1$
\\
$.GK(v_1,v_2,Jatf,Jitf)$ &
$+ katfd_{pp}.CycD0.[Mass]$ &
\\
&
$v_2=kitf_p + kitfa_{pp}.$ &

\\
&
$.[CycA]+kitfb_{pp}.[CycB]$ &

\\
\hline
\end{longtable}
}

{\small
\begin{longtable}{ p{0.20\linewidth}  p{0.75\linewidth} }
\hline
control & value \\
\hline \\
$s_1$ &
$w_{1,1}.[CycA] + w_{1,2}.[CycB] + w_{1,4}.[APCP] + w_{1,6}.[Cdc20i] + w_{1,7}.[Cdh1] + w_{1,8}.[CKI] + w_{1,9}.[Mass] - 1 > 0$\\
\hline
$s_2$ &
$w_{2,3}.[CycE] + w_{2,9}.[Mass] + w_{2,12}.[TriE] - 1 > 0$\\
\hline
$s_3$ &
$w_{3,3}.[CycE] + w_{3,9}.[Mass] + w_{3,12}.[TriE] - 1 > 0$\\
\hline
$s_4$ &
$w_{2,2}.[CycB] + w_{2,9}.[Mass] + w_{2,10}.[pB] - 1 > 0$\\
\hline
\end{longtable}
}

\subsection*{Table 2.4 - Parameters of the hybridized mammalian cell cycle
model described in the table~2.3.}

{\small
\begin{longtable}{  p{0.45\linewidth}  p{0.5\linewidth}  }
\hline
constant & value   \\
\hline \\
$ksa_{pp}$ & 0.024064 \\
\hline
$kse_{pp}$ & 0.18569 \\
\hline
$kwee_pp$ & 0.17326 \\
\hline
$k25_{pp}$ & 3.5168 \\
\hline
$ksb_{pp}$ & 0.030148 \\
\hline
$w_{1,1}$ & 1.e+9 \\
\hline
$w_{1,2}$ & 0.4352e+9 \\
\hline
$w_{1,4}$	& -1.5677e+9 \\
\hline
$w_{1,6}$ & -4.0592e+9 \\
\hline
$w_{1,7}$ & 1.e+9 \\
\hline
$w_{1,8}$ & -0.7937e+9 \\
\hline
$w_{1,9}$ & 0.1138e+9 \\
\hline
$w_{2,3}$ & -2.218e+9 \\
\hline
$w_{2,9}$ & 1.e+9 \\
\hline
$w_{2,12}$ & -10.027e+9 \\
\hline
$w_{3,3}$ & 0.2278e+9 \\
\hline
$w_{3,9}$ & -0.1015e+9 \\
\hline
$w_{3,12}$ & 1.e+9 \\
\hline
$w_{4,2}$ & 0.2294e+9 \\
\hline
$w_{4,9}$ & -0.0294e+9 \\
\hline
$w_{4,10}$ & 1e+9 \\
\hline
\end{longtable}
}

\nocite{*}
\bibliographystyle{eptcs}
\bibliography{./hsb_2013}
\end{document}